\documentclass[sigconf]{acmart}

\usepackage{xcolor}
\definecolor{DarkGreen}{HTML}{000000}

\usepackage{graphicx}
\usepackage{caption}
\usepackage{subcaption}
\usepackage{todonotes}
\usepackage{multirow}
\usepackage{colortbl}
\definecolor{lightgray}{gray}{0.9}
\AtBeginDocument{%
  \providecommand\BibTeX{{%
    \normalfont B\kern-0.5em{\scshape i\kern-0.25em b}\kern-0.8em\TeX}}}

\copyrightyear{2025}
\acmYear{2025}
\setcopyright{none}
\setcctype{by}
\acmConference[CHI '25 Workshop on Tools for Thought]{Workshop on Tools for Thought}{April 26, 2025}{Yokohama, Japan}
\acmISBN{}

\makeatletter
\renewcommand\@formatdoi[1]{\ignorespaces}

\begin{document}

\title{Supporting Students' Reading and Cognition with AI}

\begin{abstract}
With the rapid adoption of AI tools in learning contexts, it is vital to understand how these systems shape users' reading processes and cognitive engagement. We collected and analyzed text from 124 sessions with AI tools, in which students used these tools to support them as they read assigned readings for an undergraduate course. We categorized participants' prompts to AI according to Bloom’s Taxonomy of educational objectives---Remembering, Understanding, Applying, Analyzing, Evaluating. Our results show that ``Analyzing'' and ``Evaluating'' are more prevalent in users' second and third prompts within a single usage session, suggesting a shift toward higher-order thinking. However, in reviewing users' engagement with AI tools over several weeks, we found that users converge toward passive reading engagement over time. Based on these results, we propose design implications for future AI reading-support systems, including structured scaffolds for lower-level cognitive tasks (e.g., recalling terms) and proactive prompts that encourage higher-order thinking (e.g., analyzing, applying, evaluating). Additionally, we advocate for adaptive, human-in-the-loop features that allow students and instructors to tailor their reading experiences with AI, balancing efficiency with enriched cognitive engagement. Our paper expands the dialogue on integrating AI into academic reading, highlighting both its potential benefits and challenges.

\end{abstract}


\author{Yue Fu}
\email{chrisfu@uw.edu}
\orcid{0000-0001-5828-5932}
\affiliation{%
  \department{Information School}
  \institution{University of Washington}
  \city{Seattle}
  \state{Washington}
  \country{USA}
}

\author{Alexis Hiniker}
\email{alexisr@uw.edu}
\orcid{0000-0003-1607-0778}
\affiliation{%
  \department{Information School}
  \institution{University of Washington}
  \city{Seattle}
  \state{Washington}
  \country{USA}
}

\maketitle

\section{Introduction}
Reading is a cornerstone of human civilization and cognitive development, shaping the reader’s understanding of the world and themselves. The act of reading stimulates working memory, inference-making, and schema activation, which are fundamental components of comprehension and critical thinking \cite{Kintsch1978TowardAM}. It enhances vocabulary, communication, imagination, and knowledge acquisition \cite{sevensprings}. According to a 2024 global AI student survey, 86\% of students use AI tools to support their studies, highlighting the need to understand how these emerging AI tools influence users' reading and cognition \cite{survey}.

However, emerging research on AI-assisted reading and learning is mixed. On the one hand, using AI has the benefits of providing personalized learning experiences, simplifying the comprehension of complex texts and enhancing students’ overall reading effectiveness \cite{zheng2024exploration, ademola2024reading}. On the other hand, relying on AI for immediate comprehension may encourage a passive approach to reading, where students prioritize convenience over critical evaluation \cite{cahyani2023ai, ademola2024reading}. Over-reliance on AI tools could undermine deeper cognitive engagement with texts. Furthermore, the influence of AI varies by students’ baseline abilities: while lower-performing students often benefit significantly, higher-performing students may experience negative effects \cite{etkin2024differential}. Overall, empirical evidence on using AI to support reading is in its early stages.

With the widespread adoption of AI tools in both higher education and K–12 settings, it is critical to understand how these technologies influence students' learning. In this paper, we contribute to this discussion by examining how students utilize AI tools to support their reading and cognition. We conducted a mixed-methods study in an upper-level undergraduate course where students opted to use AI tools to support them as they read readings. The reading materials included academic literature, business reports, and news articles; all were typical of material assigned in a college-level class. Based on our findings, we offer design recommendations for educators and AI developers to better support students’ learning and cognitive development.

\section{Research Context and Approach
}
We collected data during an upper-level college AI class. The class introduced students to various AI application topics such as AI for science, AI for work, co-creation with AI, etc. Students could either write a 200–300 word response for each assigned reading or use any AI tool(s) to assist with their reading (we observed students mostly used ChatGPT, Gemini, Perplexity, and Claude). All students chose the AI-assisted option. Students who chose the AI-assisted option were asked to submit their logs of prompts to AI, AI responses, their reflections about using AI and the reading material, and proposed discussion questions that they came up with after their conversation with the AI tool. We analyzed the first three weeks of reading logs from which we collected 143 entries from nineteen students.
To explore how students engaged with AI, we examined both their prompts and the reflective responses they provided. We categorized each prompt according to Bloom’s Taxonomy of educational objectives (Remembering, Understanding, Applying, Analyzing, Evaluating, Creating) \cite{bloom1956taxonomy, huitt2011bloom} (see Table \ref{tab:BloomCognition} for definitions of each category). Bloom's taxonomy is a widely used framework for classifying cognitive skills and has been used to code users' cognitive involvement with AI in the prior literature \cite{lee2025the, hui2024incorporating}. This coding enabled us to detect patterns in students’ cognitive engagement with the material. By mapping AI interactions to specific categories, we gained insights into how AI might enhance or hinder different dimensions of the reading and cognitive processing.

\begin{table*}[ht]
\centering
\begin{tabular}{@{}l p{0.6\textwidth}@{}}
\toprule
\textbf{Cognition Category} & \textbf{Coding Definition} \\
\midrule
Remembering   & Recalling or recognizing information, facts, concepts, terms, or ideas. \\
\hline
Understanding & Grasping the meaning of information by explaining or summarizing. \\
\hline
Applying      & Using knowledge or techniques to solve problems in new situations. \\
\hline
Analyzing     & Examining motives, drawing inferences, finding evidence to support
                generalizations, comparing and contrasting, and detecting underlying
                themes or relationships. \\
\hline
Evaluating    & Making judgments or critique based on criteria, standards, or evidence. \\
\bottomrule
\end{tabular}
\caption{Cognitive Activities based on Bloom's taxonomy \cite{bloom1956taxonomy}}
\label{tab:BloomCognition}
\end{table*}

\section{Results}
\begin{figure}[h]
    \centering
    \includegraphics[width=0.5\textwidth]{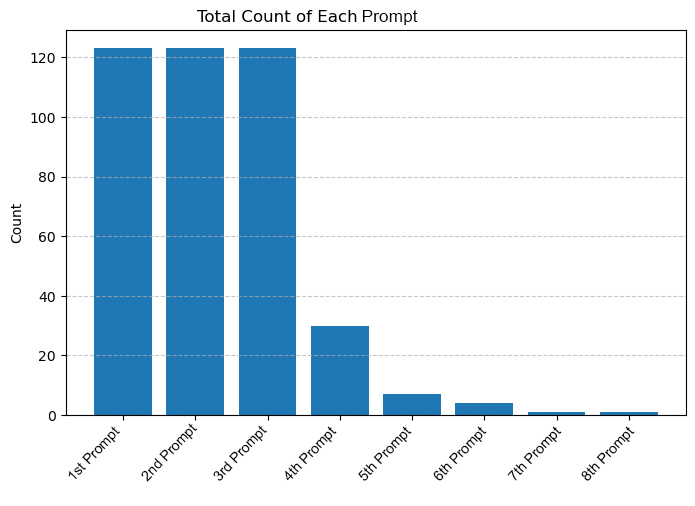}
    \caption{Total Count of Each Prompt for All Readings}
    \label{fig:prompt_count}
\end{figure}

We collected 143 reading entries with multiple prompts to the AI tool for each entry. Entries from three participants (P1, P4, P20) used exact material from the reading task assignment info sheet to prompt the AI for all reading assignments, thus we decided to exclude these participants from our final analysis. In total, we coded prompts and responses from 124 reading entries. The average number of prompts per participant per reading was 3.35. Students were required to prompt the AI tool at least three times (leading to 3 x 124 prompts for each entry); additional prompting declined significantly and participants produced: 31 fourth prompts, 7 fifth prompts, 4 sixth prompt, 1 seventh prompt and 1 eighth prompt. This pattern suggests that students rarely explored beyond the required interactions, indicating limited iterative engagement with AI to support their reading (see Figure \ref{fig:prompt_count}).

\begin{figure}[h]
    \centering
    \includegraphics[width=0.5\textwidth]{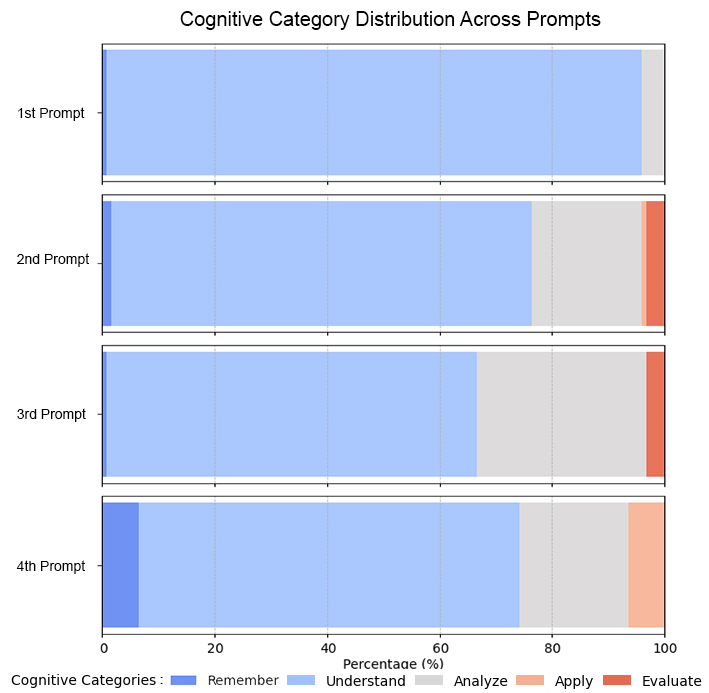}
    \caption{Cognitive Category Distribution Across 1st to 4th Prompt}
    \label{fig:prompt_distribution}
\end{figure}

\subsection{Differences in Cognitive Category Distribution Across Prompt Sequence}

The majority of prompts fell under the ``Understanding'' category, which aligns with the reading-based nature of the task (see Figure \ref{fig:prompt_distribution}). However, we observed a shift in cognitive complexity across prompt stages. In the second and third prompts, students increasingly engaged in ``Analyzing'' and ``Evaluating'' categories, indicating a move toward higher-order thinking. Interestingly, in the fourth prompt, students returned to ``Understanding'' based queries, suggesting a tendency to summarize or consolidate key takeaways before concluding their interaction with AI. This pattern implies that students may use AI to reinforce comprehension at the final stage of engagement. The findings show that different prompts belong to varying levels of cognitive processing, emphasizing how AI interactions change within a prompting sequence.

\subsection{Differences in Cognitive Category Distribution Across Weeks}
\begin{figure}[h]
    \centering
    \includegraphics[width=0.5\textwidth]{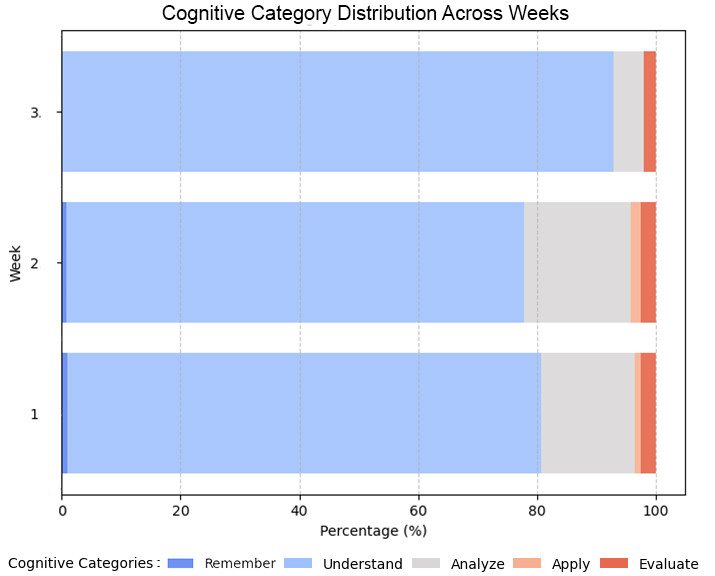}
    \caption{Cognitive Category Distribution Across Weeks}
    \label{fig:weeks_distribution}
\end{figure}
In Weeks 1 and 2, approximately 80\% of prompts fell into the ``Understanding'' category, reflecting a strong focus on comprehension (see Figure \ref{fig:weeks_distribution}). In Week 3, the proportion of ``Understanding'' prompts increased further to 92.86\%, indicating a shift toward more passive reading engagement. This suggests that students' cognitive engagement patterns changed over time, with a possible decline in the willingness to use higher-order thinking as students become familiar with using AI to support their reading comprehension. An alternative explanation is that students may have begun the semester with greater motivation, which gradually diminished as academic demands accumulated, leading to a preference for quicker, more surface-level interactions with AI tools.

\subsection{Cognitive Category Distribution Students’ Submitted Discussion Question}
\begin{figure}[h]
    \centering
    \includegraphics[width=0.5\textwidth]{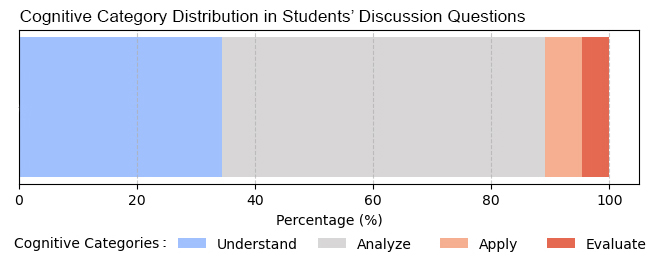}
    \caption{Cognitive Category Distribution in Students' Submitted Discussion Question}
    \label{fig:question_distribution}
\end{figure}
Each student also submitted 1-3 discussion questions per reading, some of which fell into multiple cognitive categories (e.g., ``Understanding'' and ``Analyzing'') (see Figure \ref{fig:question_distribution}). In subsequent analysis, we coded these entries to calculate the percentage of each cognitive category. The majority of discussion questions fell into the ``Analyzing'' category (54.69\%), followed by ``Understanding'' (34.38\%), ``Applying'' (6.52\%), and ``Evaluating'' (4.68\%).

\subsection{Students' Qualitative Experiences and Reflections on Using AI to Support Reading}
\label{qualitative results}

Students overwhelmingly expressed positive sentiments about using AI to support reading. Their reflections highlighted the following perceived benefits of AI assistance for their reading:
\begin{itemize}
    \item Increased reading efficiency
    \item Concise and structured summaries
    \item Support in understanding difficult terminologies
    \item Encouragement of critical thinking
    \item Digestible formats such as bullet points
    \item Connection to external resources beyond the assigned reading
    \item Generation of potential follow-up questions at the end of responses
\end{itemize}
Regarding what worked well in their interaction process, students identified the following effective strategies for engaging with AI:
\begin{itemize}
    \item Comparing and contrasting varied ideas to critically analyze the author’s viewpoint
    \item Asking follow-up questions to deepen understanding
    \item Requesting explanations using specific examples and analogies
    \item Breaking down complex questions into sub-questions and prompting sequentially
    \item Summarizing the reading first and then pinpointing specific sections of interest
\end{itemize}

These qualitative insights suggest that students found AI to be a valuable tool in supporting comprehension, critical thinking, and engagement with reading materials.

\section{Discussion}

Our results indicate that students generally have a positive experience using AI to support reading. Most prompts are oriented toward the ``Understanding'' category of Bloom’s Taxonomy, reflecting a primary focus on summarization. This pattern is unsurprising given the assignment requirement to grasp essential reading content. We also see during the second and third prompt cycles show a noticeable rise in ``Analyzing'' and ``Evaluating.'' This progression indicates that once students have established a baseline comprehension, some naturally explore deeper cognitive engagement. In these prompts, students ask questions requiring inference or generalization beyond the reading text. In addition, students’ submitted discussion questions at the end of each reading assignment are also dominated by ``Analyzing,'' suggesting that when students craft open-ended inquiries for class discussion, they often probe beyond surface-level details. They ask about inferences not explicitly covered in the readings, explore cross-domain comparisons, and probe underlying relationships. This pattern points to AI’s potential to foster more cognitively challenging questions and deeper engagement \cite{hui2024incorporating}.

On the other hand, despite early signs of higher-order engagement, the data reveal a growing reliance on ``Understanding'' prompts by Week 3 compared to previous weeks. Students appear to streamline their approach, possibly imitating previous prompt patterns rather than experimenting with new ways to use AI. Moreover, although many students mention the importance of iterating with follow-up questions (as shown in Section \ref{qualitative results}), few actually surpass four prompt interactions. This discrepancy between students’ self-reported strategies in the qualitative data and actual usage could point to time constraints, motivational factors, or simply habit-formation in using AI primarily for quick comprehension rather than iterative deep inquiry.

Collectively, these findings underscore both the promise and the pitfalls of AI-assisted reading. On one hand, AI can help students quickly comprehend texts and guide them toward more analytical or evaluative questions. On the other hand, without targeted scaffolding and deliberate design features, students may default to superficial uses of AI. Over-reliance on summarization tools can foster passive reading behaviors \cite{cahyani2023ai}.

AI systems can shape human cognitive behavior through their design, responses, and interaction patterns. How information are structured and presented can encourage engagement of certain cognitive responses. For instance, if an AI system offers prompts that explicitly invite deeper reflection (e.g., ``Would you like to compare multiple viewpoints or examine underlying causes?''), it could encourage students to analyze and evaluate rather than just memorize. Conversely, passive or AI responses can reinforce surface-level comprehension as the end goal. How to balance there forces is the key to support students' reading and cognition. Here we offer our suggestions for system designers to consider:

\vspace{3mm}

\textbf{\textit{Support Low-Level Cognitive Skills:}}

\begin{itemize}
    \item \textbf{Fostering Recall and Memorization:} Systems can provide fact sheets, concept lists, or ``cue cards'' at the outset or conclusion of each session. These resources can help learners remember key terms and ideas, supporting future recall.
    \item \textbf{Facilitating Nuanced Understanding:} Rather than presenting a single ``summary,'' AI tools can highlight specific sections—methods, findings, or discussions—to help students pinpoint areas they find challenging. A ``zoom in'' and ``zoom out'' approach (e.g., concept maps) could prevent premature convergence \cite{sarkar2023exploring} on a single aspect of the reading.
\end{itemize}

\textit{\textbf{Support Higher-Order Cognitive Skills:}
}
\begin{itemize}
    \item \textbf{Prompting Analysis and Evaluation:} AI systems can intentionally steer conversations toward deeper thinking by offering optional follow-up questions focused on analyzing, applying, or evaluating.
    \item \textbf{Facilitating Application Across Contexts:} Students often struggle to connect reading insights to other domains. Systems can scaffold this transfer by prompting learners to link newly acquired knowledge to practical examples or emerging research.
    \item \textbf{Promoting Metacognitive Reflection \cite{tankelevitch2024metacognitive}:} Features like quizzes, flashcards, or reflective prompts can reinforce content retention and encourage students to articulate what they have learned and how they arrived at that understanding. Designed delay and pause could be used \cite{buccinca2021trust, fu2025exploring}. For example, a system can require users to spend certain amount of time crafting their prompts with the help and suggestion from the system.
\end{itemize}

\textit{\textbf{Balance Proactive and Adaptive Features:}
}
\begin{itemize}
    \item An ideal AI reading partner would adapt its prompts and scaffolding to the user’s evolving goals and skill level. Proactive systems can be beneficial if they encourage proper inquiry. However, overstepping user autonomy risks replicating the Microsoft ``Clippy'' problem \cite{cain2017life}---an intrusive system that frustrates rather than assists. A balanced approach involves offering guidance without overwhelming or constraining the student.
\end{itemize}

\textbf{\textit{Emphasize a Human-in-the-Loop Approach:}}
\begin{itemize}
    \item \textbf{Customizable Scaffolding:} Because predicting user intentions is complex, giving learners (and instructors) control over the AI’s ``level of guidance'' is essential. Through adjustable settings, students can specify their desired reading goals, depth of analysis, language level, or type of feedback they want from the system.

    \item \textbf{Teacher as Facilitator:} By configuring AI tools, instructors can align the system's scaffolding with specific course objectives, ensuring that reading support complements broader learning outcomes. Teachers can also adapt AI-driven experiences to accommodate diverse student needs and specific learning goals of reading materials.

\end{itemize}

\bibliographystyle{ACM-Reference-Format}
\bibliography{references}

\end{document}